\documentclass[twocolumn,trackchanges]{aastex62} 

\graphicspath{{./}{figures/}}

\shorttitle{Galactic disk asymmetry between 12$-$15 \,kpc}
\shortauthors{Wang et al.}

\begin{document}

\title{Mapping the Galactic disk with the LAMOST and Gaia Red clump sample: III: A new velocity substructure and time stamps of the Galactic disk asymmetry in the disk between 12$-$15 \,kpc}
\author[0000-0001-8459-1036]{Hai-Feng Wang}
\affil{South$-$Western Institute for Astronomy  Research, Yunnan University, Kunming, 650500, P.\,R.\,China}
\affil{Department of Astronomy, China West Normal University, Nanchong 637009, China}
\affil{LAMOST Fellow }
\author{Jeffrey L. Carlin}
\affil{LSST, 950 North Cherry Avenue, Tucson, AZ 85719, USA}
\author{ Y. Huang}
\affil{South$-$Western Institute for Astronomy  Research, Yunnan University, Kunming, 650500, P.\,R.\,China}
\author{Mart\'\i n L\'opez-Corredoira}
\affil{Instituto de Astrof\'\i sica de Canarias, E-38205 La Laguna, Tenerife, Spain}
\affil{Departamento de Astrof\'\i sica, Universidad de La Laguna, E-38206 La Laguna, Tenerife, Spain}
\author{ B.-Q. Chen}
\affil{South$-$Western Institute for Astronomy  Research, Yunnan University, Kunming, 650500, P.\,R.\,China}
\author{ C. Wang}
\affil{Department of Astronomy, Peking University, Beijing 100871, People's Republic of China.}
\affil{LAMOST Fellow }
\affil{Kavli Institute for Astronomy and Astrophysics, Peking University, Beijing 100871, People's Republic of China}
\author{ J. Chang}
\affil{Key Laboratory of Optical Astronomy, National Astronomical Observatories, Chinese Academy of Sciences, Beijing 100012, People's Republic of China}
\affil{Purple Mountain Observatory, the Partner Group of MPI f$\ddot{u}$r Astronomie, 2 West Beijing Road, Nanjing 210008, China}
\author{ H.-W. Zhang}
\affil{Department of Astronomy, Peking University, Beijing 100871, People's Republic of China.}
\affil{Kavli Institute for Astronomy and Astrophysics, Peking University, Beijing 100871, People's Republic of China}
\author{ M.-S. Xiang}
\affil{Max-Planck Institute for Astronomy, Konigstuhl, D-69117, Heidelberg, Germany}
\affil{Key Laboratory of Optical Astronomy, National Astronomical Observatories, Chinese Academy of Sciences, Beijing 100012, People's Republic of China}
\author{ H.-B. Yuan}
\affil{Department of Astronomy, Beijing Normal University, Beijing 100875, People's Republic of China}
\author{ W.-X. Sun}
\affil{South$-$Western Institute for Astronomy  Research, Yunnan University, Kunming, 650500, P.\,R.\,China}
\author{ X.-Y. Li}
\affil{South$-$Western Institute for Astronomy  Research, Yunnan University, Kunming, 650500, P.\,R.\,China}
\author{ Y. Yang}
\affil{South$-$Western Institute for Astronomy  Research, Yunnan University, Kunming, 650500, P.\,R.\,China}
\author{ L.-C. Deng}
\affil{Key Laboratory of Optical Astronomy, National Astronomical Observatories, Chinese Academy of Sciences, Beijing 100012, People's Republic of China}
\affil{Department of Astronomy, China West Normal University, Nanchong 637009, China}
\correspondingauthor{HFW; YH}
\email{ hfwang@bao.ac.cn{(\rm HFW)};\\
yanghuang@pku.edu.cn{\rm (YH)}}; \\

\begin{abstract}
We investigate the three-dimensional asymmetrical kinematics and present time stamps of the Milky Way disk between Galactocentric distances of $R=12$ and 15 \,kpc, using red clump stars selected from the LAMOST Galactic survey, also with proper motion measurements provided by the Gaia DR2. We discover velocity substructure above the Galactic plane corresponding to a density dip found recently (``south-middle opposite'' density structure[R $\sim$ 12-15 \,kpc, Z $\sim$ 1.5 \,kpc] discovered in \citet{Wang2018b, Wang2018c}) in the radial and azimuthal velocity. For the vertical velocity, we detect clear vertical bulk motions or bending mode motions, which has no clear north-south asymmetry corresponding to the in-plane asymmetrical features. In the subsample of stars with different ages, we find that there is little temporal evolution of the in-plane asymmetry from 0$-$14 \,Gyr, which means the structure is sensitive to the perturbations in almost cosmic time possibly.  We propose that the possible scenario of this asymmetric velocity structure is caused by the mechanisms generated in-plane, rather than vertical perturbations.
\end{abstract}

\section{Introduction} 

A basic assumption often used to interpret observations is that galaxies are in equilibrium or stationary in the potential. This was shown to be an invalid assumption with the evidence for a Galactic North-South asymmetry in the number density and bulk velocity of solar neighborhood stars revealed by \citet{Widrow12}. These observations began what is known as Galactoseismology for the Milky Way. Our home galaxy disk is a typical dynamical system that is perturbed by bars, giant molecular clouds, spiral structures, warps, tidally disrupting satellite galaxies, and dark matter subhalos. Some imprints will be left on stars by these processes \citep{Widrow12}. More recently, \citet{antoja2018} found that the disk is full of velocity substructures in phase space caused by phase mixing, showing that modelling the Galactic disk as a time-independent axisymmetric component is definitively incorrect. We are entering into the golden era of galactoseismology with the help of Gaia data \citep{Gaia2018}.

Disk oscillations out to 25 \,kpc were unveiled by \citet{xu2015}. The wave-like pattern with the help of star counts shows four overdensities: in the north at distances of about 2 kpc (North near), in the south at 4$-$6 kpc (South middle), in the north at distances of 8$-$10 kpc, and in the south at distances of 12$-$16 \,kpc from the Sun in the anticenter direction. \citet{Wang2018b, Wang2018c} detected rich substructures displayed in the residuals of stellar density. Among them, the substructures O14$-$1.5 and D14 + 2.0 show a north-south asymmetry, the first one corresponds to the south$-$middle structure with high confidence, and the second one corresponds to a new substructure located in the opposite side of the south middle region (named as ``south middle opposite"), of which locations are  R = 8.5 to 15 \,kpc, Z = -1 to -2.5 \,kpc and R = 12 to 15 \,kpc, Z$\sim$1.5 \,kpc, separately. 

Followed by this, \citet{Wang2018a} mapped the north near region by kinematical analysis with significant larger error than Gaia data. Before Gaia era, most of these asymmetrical works are only shown in density and  a few works shown in kinematics with lower precision of proper motion. Now, we have unprecedented proper motion with the help of Gaia and millions of stars with age from LAMOST \citep{Deng2012, liu2014, cui2012, zhao2012} so that we can constrain the kinematics and time evolution of these asymmetrical structures.

Recently, time stamps and dynamical analysis have been placed on the first north near density structure around 11 \,kpc \citep{wang2019b, wang2019c}. In this work, we propose to shift our horizon to the south middle or opposite region around 12$-$15\,kpc with red clump stars mainly in the anticenter. 
We take advantage of age determinations for red clump stars derived by \citet{Huang2019}, by which we can track the velocity temporal evolution of different stellar populations. 
The combination of Gaia and LAMOST provides us a good sample to study kinematics in the disk.

This paper is organized as follows. In Section 2, we introduce how we select the red clump star sample and describe its properties. The results and discussions are presented in Section 3. Finally, we conclude this work in Section 4.

\section{The Sample Selection}  

A sample of over 150,000 primary Red Clump (RC) stars from the LAMOST Galactic spectroscopic surveys \citep{Deng2012, liu2014, cui2012, zhao2012} is selected based on their positions in the metallicity-dependent effective temperature--surface gravity and color--metallicity diagrams with the help of asteroseismology data. Thanks to the  LAMOST spectra and the Kernel Principal Component Analysis (KPCA) method, the uncertainties of distances for our primary red clump stars are no more than 5 to 10 percent, and the uncertainties on stellar masses and ages are 15 and 30 percent, respectively \citep{Huang2019}.

The Gaia survey aims to measure the astrometric, photometric, and spectroscopic parameters of 1\% of the stellar population in the Milky Way \citep{Gaia2016}. The recently released Gaia DR2 provides a catalog of over 1.3 billion sources \citep{Gaia2018} with unprecedented high-precision proper motions with typical uncertainties of 0.05, 0.2 and 1.2 mas yr$^{-1} $ for stars with G-band magnitudes $\leq$ 14, 17 and 20 \,mag, respectively. The solar motion we adopt is that of \citet{Tian15}: [$U_{\odot}$ $V_{\odot}$ $W_{\odot}$] = [9.58, 10.52, 7.01] km s$^{-1} $. The circular speed of the Local Standard of Rest is adopted as 238 km s$^{-1} $ \citep{Schonrich12}. Radial velocities of the samples are corrected by adding 3.1 km s$^{-1}$ based on comparisons with APOGEE data \citep{Blanton17}. We find that our results are stable even if we choose other solar motions determined by \citet{Huang2015}.

In order to build the sample containing stellar astrophysical parameters and precise kinematical information, we combine the LAMOST spectroscopic survey and Gaia catalogs; we remove stars with duplicated observations and without parameters such as distance, radial velocity, temperature and surface gravity.  We only select stars located inside $|Z|$ $<$ 3 kpc and 12 $<$ R $<$ 15 \,kpc. The stars with LAMOST spectroscopic SNR $<$ 20 and age larger than 15 \,Gyr are not included. We have also checked whether or not we should set criteria for the relative error of age larger than 50$\%$ in this work, the results are stable. In order to reduce the contamination of halo stars but try to keep as many stars as possible in the outer disk, we only use stars with [Fe/H] larger than $-$1.5 \,dex and we set some criterias in velocity to remove fast-moving halo stars: $V_R$=[-150, 150] km s$^{-1}$, $V_{\theta}$=[-50, 350] km s$^{-1}$, and $V_Z$=[-150, 150] km s$^{-1}$.

\begin{figure*}
  \centering
  \includegraphics[width=0.7\textwidth]{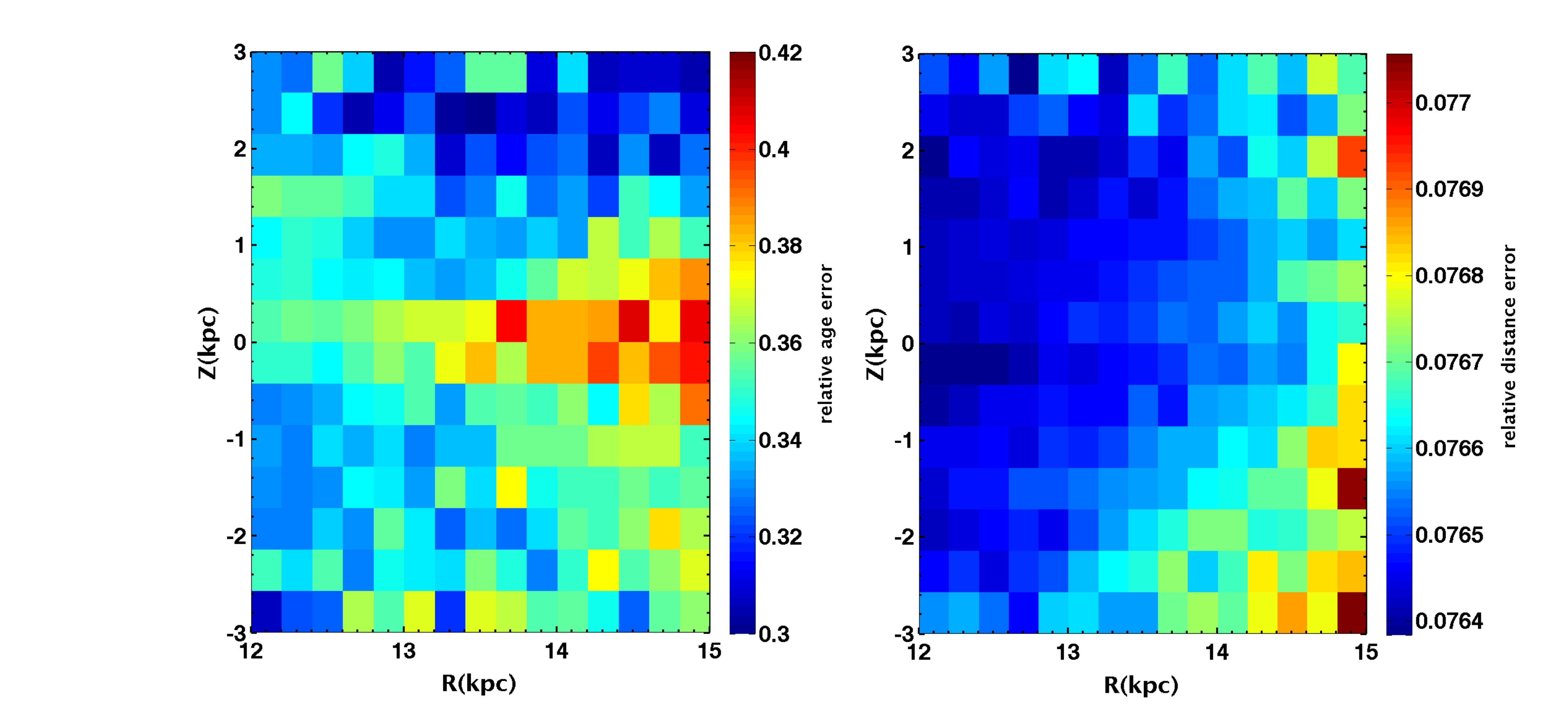}
  \caption{Left and right panels show the distribution in the R$-$Z plane of projected relative age and distance errors, respectively. The color bars indicate the relative age errors (left) and relative distance errors (right). The blue/cyan color dominate the pattern implying that most of the relative age error is less than 38 \% and most of the relative distance error is less than 8 \%.}
  \label{xyz_12}
\end{figure*}

The spatial distributions of the sample in the Cartesian coordinate system are displayed in Fig. \ref{xyz_12}. We only focus on the R=[12 15] and Z=[-3 3] region. Left and right (R$-$Z) panels show the projected relative age error and relative distance errors distributions, respectively. The color bars indicate the relative age errors (left panel, \citet{Huang2019}), it shows that almost all stars error is in the 30\% to 40\%, the mean value is around 35\%. In the left panel we see the relative error value clearly goes in excess of 35\% for -0.5 < Z < 0.5 and R > 13.5, which might be due to that the training data set and the KPCA methods are not perfect. Red clump stars are standard candles, It is not strange that the relative distance error of our sample is less than 8\%.

\section{Results and Discussion} 

\subsection{Asymmetry in the 12$-$15 \,kpc range corresponding to the ``south-middle opposite'' density structure }

\begin{figure*}[!t]
\centering
\includegraphics[width=0.3\textwidth, trim=0.0cm 0.0cm 0.0cm 0.0cm, clip]{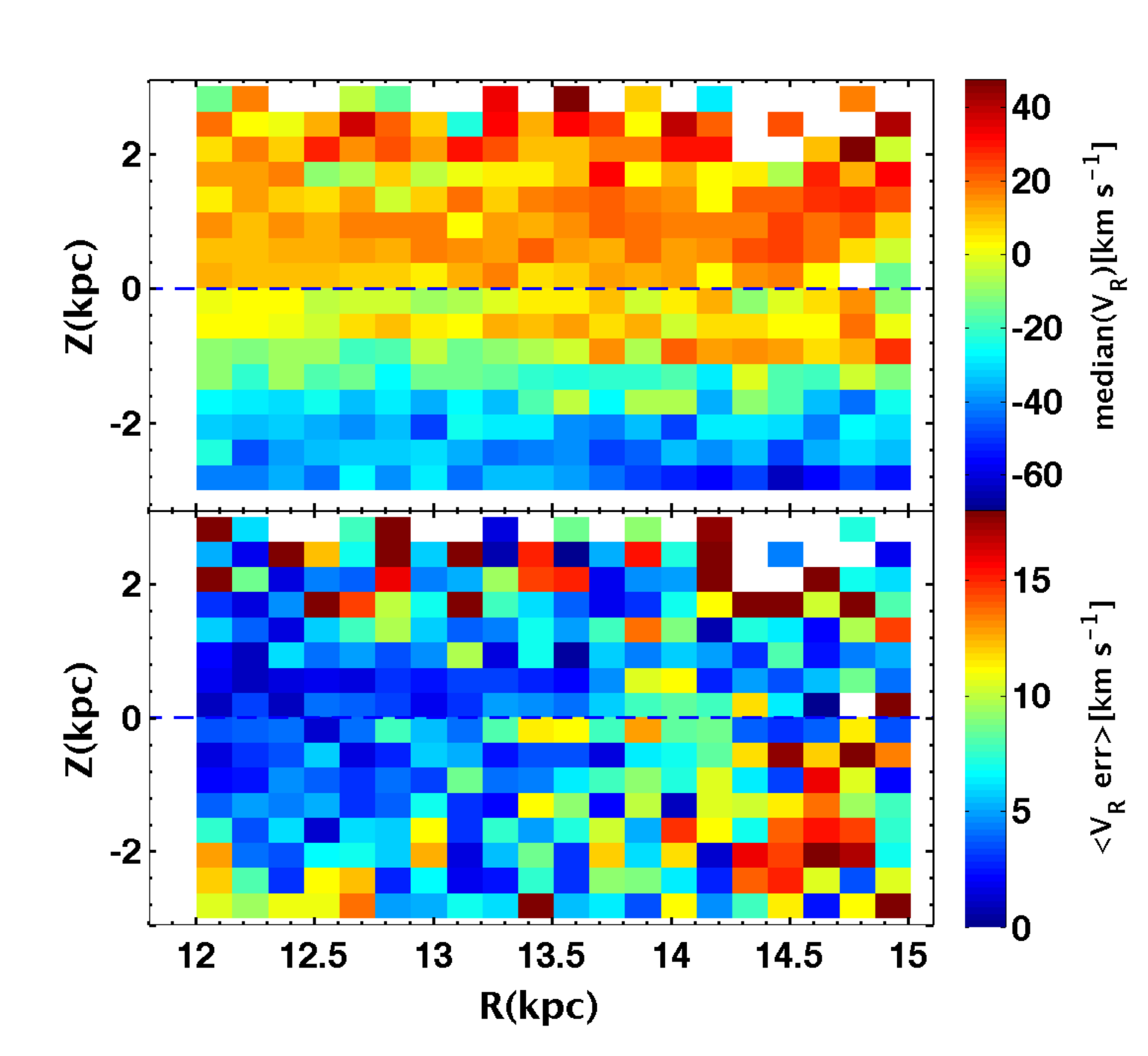}
\includegraphics[width=0.3\textwidth, trim=0.0cm 0.0cm 0.0cm 0.0cm, clip]{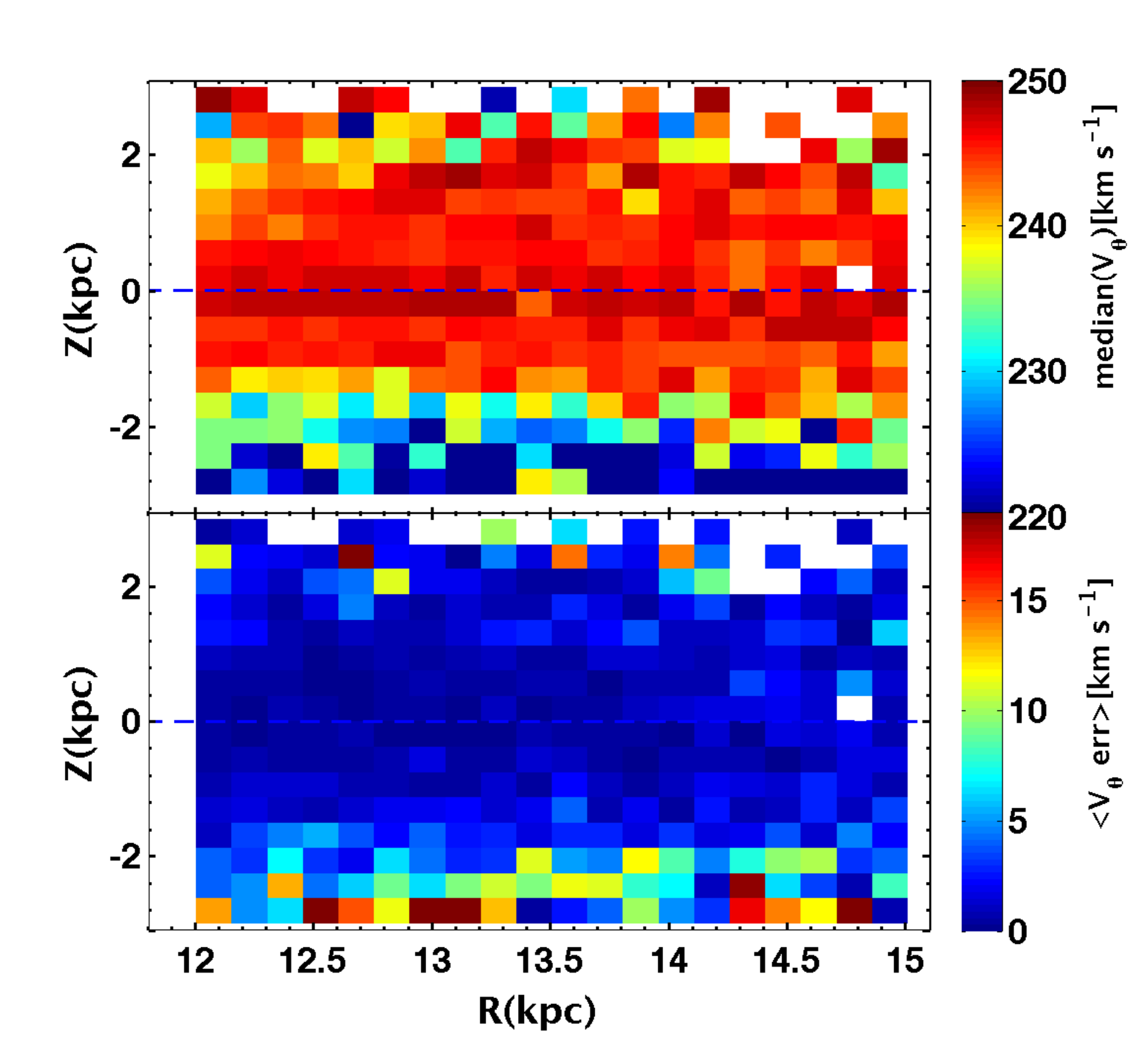}
\includegraphics[width=0.3\textwidth, trim=0.0cm 0.0cm 0.0cm 0.0cm, clip]{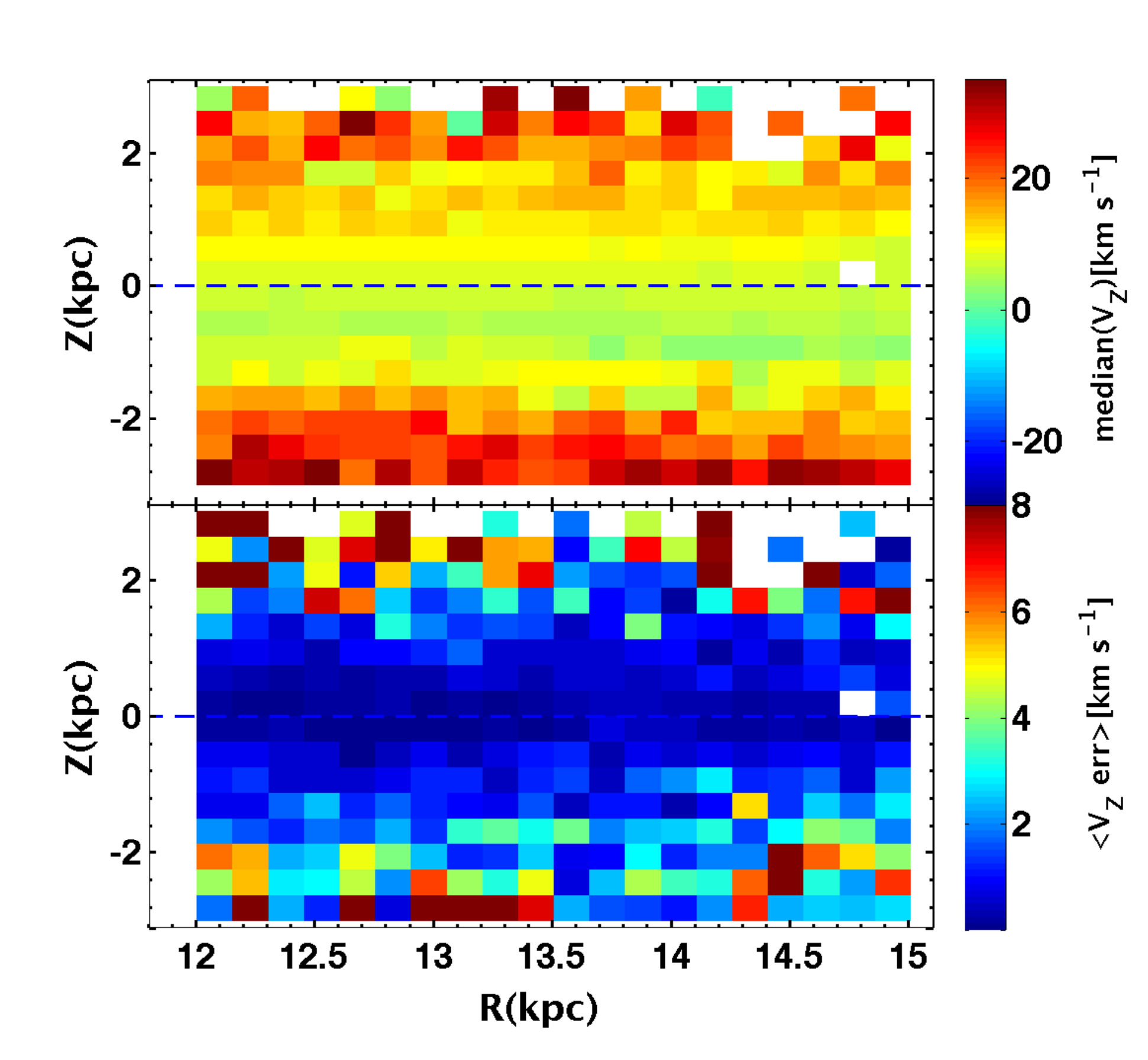}
\caption{Edge$-$on views of the kinematics of the disk derived using the red clump giant sample. From left  to right, bins show the median velocity $V_{R}$, $V_{\theta}$, $V_{Z}$ (in km~s$^{-1}$) of RC stars within each bin on the R, Z plane. The red color is prominent in the north corresponding to the blue/cyan color in the south, the  asymmetric region corresponding to the south middle and south middle opposite density structures is evident in the radial and rotational velocity distributions. The right panel shows vertical velocity distribution in the R, Z plane, the range is R=[12 15]\ ,kpc, Z=[-3 3] \,kpc, the green/red colors predominate bins both above and below the plane, illustrating that there is a clear bulk motions at all radii.} 
\label{VRVPHIVZ}
\end{figure*}

With the help of LAMOST and Gaia common stars, we show the 3D velocity distribution in Galactocentric cylindrical velocities ($V_R, V_\theta, V_Z$) on the R$-$Z plane in the range of 12$-$15 \,kpc in Fig. \ref{VRVPHIVZ}. The top left panel shows the radial velocity distribution, we can see the red colors dominate the north stars and blue/cyan colors dominate the south stars, where red colors represent positive values and blue colors represent negative velocities. There is clear asymmetry in $V_R$ between the northern and southern stars, especially for heights larger than 0.5 \,kpc. The general trend shows northern stars are moving outward and southern stars  are moving inward, showing an asymmetry from north to south. \citet{Wang2018b} found a dip in density at R = 12$-15$ \,kpc, Z $\approx$ 0.5$-$3 \,kpc (shown in Figure 3 of their work), which was described as a new substructure (we named it the ``south-middle opposite'') located in the opposite side of the south middle structure of \citet{xu2015}. This radial velocity pattern corresponds to the new substructure in kinematical space. There is a small error around a few km s$^{-1}$ for most of the bins in the bottom left panel, as determined by a bootstrap process. The high error bins at the top part of bottom panel might be caused by sampling is not very high and homogenous, the red giants contamination, stellar parameters error and random fluctuations are also possible, so the error of some bins are larger. However, if we check the colors carefully by plotting again, we find only very few bins have extremely colors or values in the bottom panel compared with the top panel. The  north$-$south absolute difference of the top one can reach around 20$-$25 km s$^{-1}$, but most of the error value is less than 5 km s$^{-1}$, only a few bins have value around 15$-$20 km s$^{-1}$. On the other hand, most stars velocity of both sides of the top panel is significantly larger than corresponding error bins in the bottom panel. Generally, we are confident that the asymmetrical signal is true by careful check and statistical analysis.

The distribution of $V_{\theta}$ (seen in the middle panel of Figure~\ref{VRVPHIVZ}) has different trends from $V_R$, exhibiting a gradient with $|$Z$|$ at south side and the northern stars gradient is weaker, and the red color is prominent in the north corresponding to the blue/cyan color in the south, which imply there are asymmetrical rotational motions when $|$Z$|$ =[0.5 2.8] \,kpc, stars in the north rotate faster than those in the south by $\sim$20$-$30 km s$^{-1}$. This kinematical structure is corresponding to the ``south-middle opposite" density structure clearly. Although it is contributed by an increase in asymmetric drift with Z as mentioned by \citet{Katz2018}, we can still suggest these asymmetries imply that the ``south-middle opposite" density structure has a larger rotational velocity than the symmetrical region. we can see the color scales, the error of most stars are less than 5 km s$^{-1}$, and very few bins are larger than 10$-$15 km s$^{-1}$, most are located in the south due to the low sampling rates or random fluctuations for this survey. When we look back to the top one, the south-north asymmetry is larger than 20-30 km s$^{-1}$ for the general trend, so for our current understanding, at least for the general pattern, our asymmetry conclusion is robust.

It is intriguing to see stars on the north and south sides of the plane at R $>$ 12 \,kpc exhibit net upward vertical motions larger than 5$-$10 km~s$^{-1}$, as shown in the right panel. The green and red colors predominate bins both above and below the plane, illustrating that regardless of position all of those stars are displaying the same vertical bulk motion.This is a typical bending mode as discussed in \citet{Widrow12, Widrow14,Chequers18}, or vertical upward bulk motions. There is no clear north$-$south vertical asymmetry and stars on both side have significant bending mode motions in the range of 12$-$15 \,kpc. Although there are some scatters in the error distribution, it is clear that most of error values is not  so large that can undermine our asymmetry conclusion at all.

\subsection{Constraining the time of the perturbation via RC ages}

\begin{figure*}
  \centering
  \includegraphics[width=0.98\textwidth]{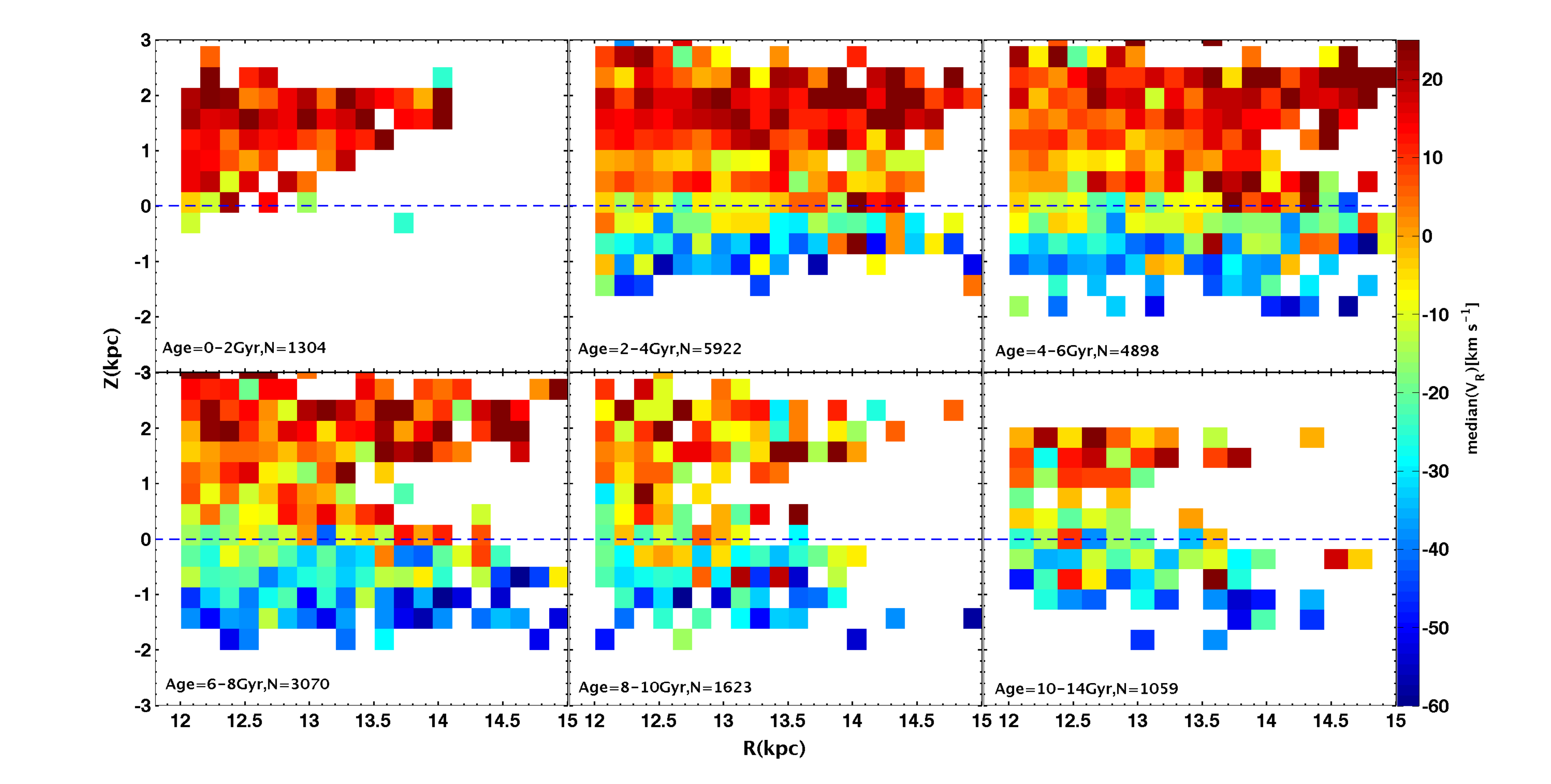}
  \caption{Radial asymmetrical structures in the R, Z plane of LAMOST-Gaia stars in different age populations. Bins in each panel are color-coded as the median radial velocity $V_R$, with the panels showing different age bins as labeled. The red/blue difference between above-the-plane and below-the plane represents  the asymmetry in radial velocity. The panels showing ages between 2-8 \,Gyr have clear north/south asymmetries in $V_R$, and these are still apparent at ages of 10$-$14 \,Gyr. Each pixel plotted in this figure has at least 5 stars and the total number is shown in each panel.}
  \label{VRage}
\end{figure*}

We divide the red clump stars sample into six stellar age ($\tau$) bins from 0 \,Gyr to 14.0 \,Gyr, to investigate the asymmetrical patterns at different stellar ages. Each stellar bin in all figures, e.g., Fig. \ref{VRage} and \ref{Vthetaage}, includes at least five stars in all panels. The total number of stars ($N$) in each age bin is labeled in all four figures of this section, we give the time stamps for describing 2D asymmetrical  radial motions and rotational motions in the Galactic disk. As we can see, there is a low sampling rate in the south side on the top left panel (0$-$2 \,Gyr bins) for all four figures. During this work, we are only interested in the sensitive time of this structure, so that the poorly-sampled regions don't influence our conclusions.

\begin{figure*}
  \centering
  \includegraphics[width=0.98\textwidth]{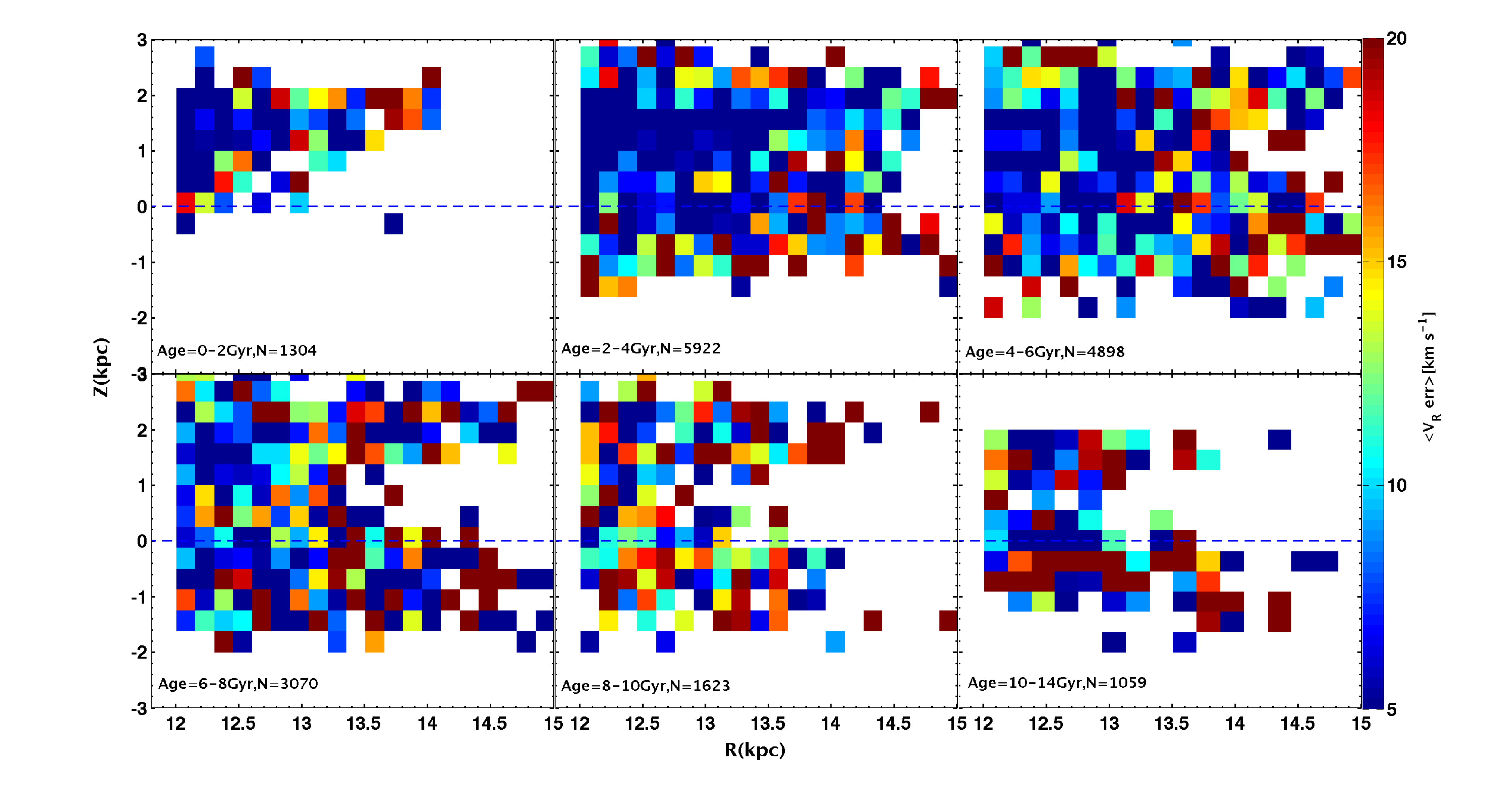}
  \caption{Error analysis of the radial asymmetrical structures in the R, Z plane for Fig.~\ref{VRage}, with each panel showing bootstrap errors for the corresponding quantities in each bins of Fig.~\ref{VRage}. Most of the value is from 5-10 km s$^{-1}$ with blue color and some values can reach 15-20 km s$^{-1}$ with red color.}
  \label{VRageerror}
\end{figure*}

\begin{figure*}
  \centering
  \includegraphics[width=0.98\textwidth]{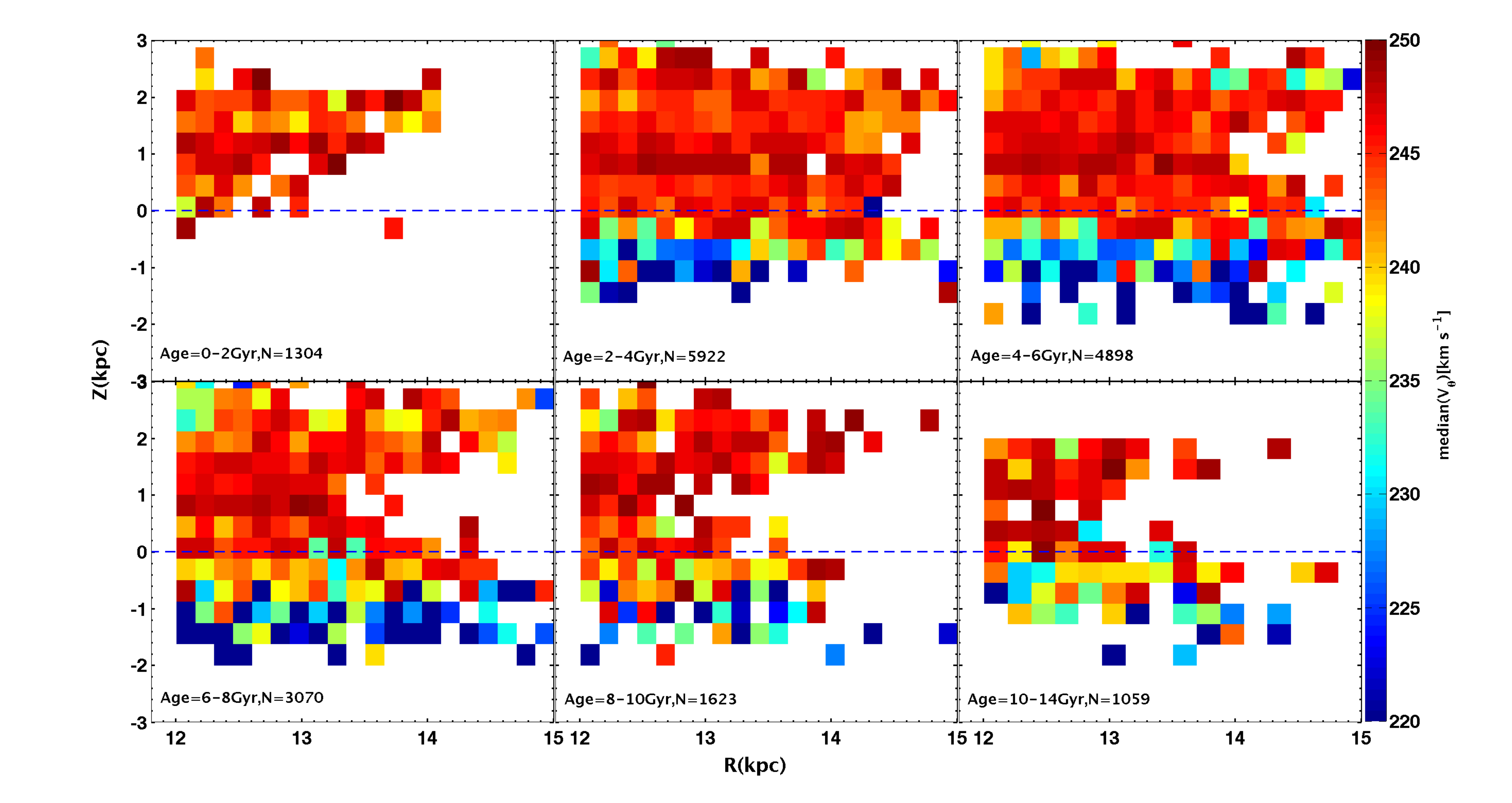}
  \caption{Similar to Fig. \ref{VRage},  but for the rotational ($V_\theta$) asymmetrical structures in the R, Z plane of the LAMOST-Gaia stars in different age populations, red color is prominent in the north with larger value. Almost every panel has asymmetries between 12$-$15 \,kpc when comparing northern and southern stars. }
  \label{Vthetaage}
\end{figure*}

The panels of Fig.~\ref{VRage} show the variation of Galactocentric radial velocity $V_R$  with age in the $R, Z$ plane, the pattern is similar with the left panel of Fig.~\ref{VRVPHIVZ} . The red/blue difference between above-the-plane and below-the plane represents a clear north/south asymmetry in radial velocity, red color means that the stars are moving outward and blue color is moving inward.  There is a large prominent velocity asymmetrical structure on the northern ($0 < Z < 3$ \,kpc) side with red bins compared to the south side in ages from 0$-$8 \,Gyr. When the age becomes older than 8 \,Gyr, there is still a weak north-south asymmetry in the range of 12$-$13 \,kpc, which implies that the ``south middle opposite'' substructure is made up of stars of all ages spanning nearly cosmic time. As we can see in Fig.~\ref{VRageerror} for the error analysis of the radial asymmetry, the error value of the most bins is around 5 km s$^{-1}$ with some redder bins, and for the age is older than 8 \,Gyr, the redder bins is more due to the age error, sampling rates or random fluctuations, stellar parameter precision, etc. Some of bins even reach around 20 km s$^{-1}$, However, we can see the clear asymmetrical pattern in Fig.~\ref{VRage} , the velocity of north stars can reach 20-25 km s$^{-1}$, south stars can reach 50-60 km s$^{-1}$, the absolute difference even can reach 30-40 km s$^{-1}$ for every age population, so we don't think the error can undermine our asymmetrical results.

\begin{figure*}
  \centering
  \includegraphics[width=0.98\textwidth]{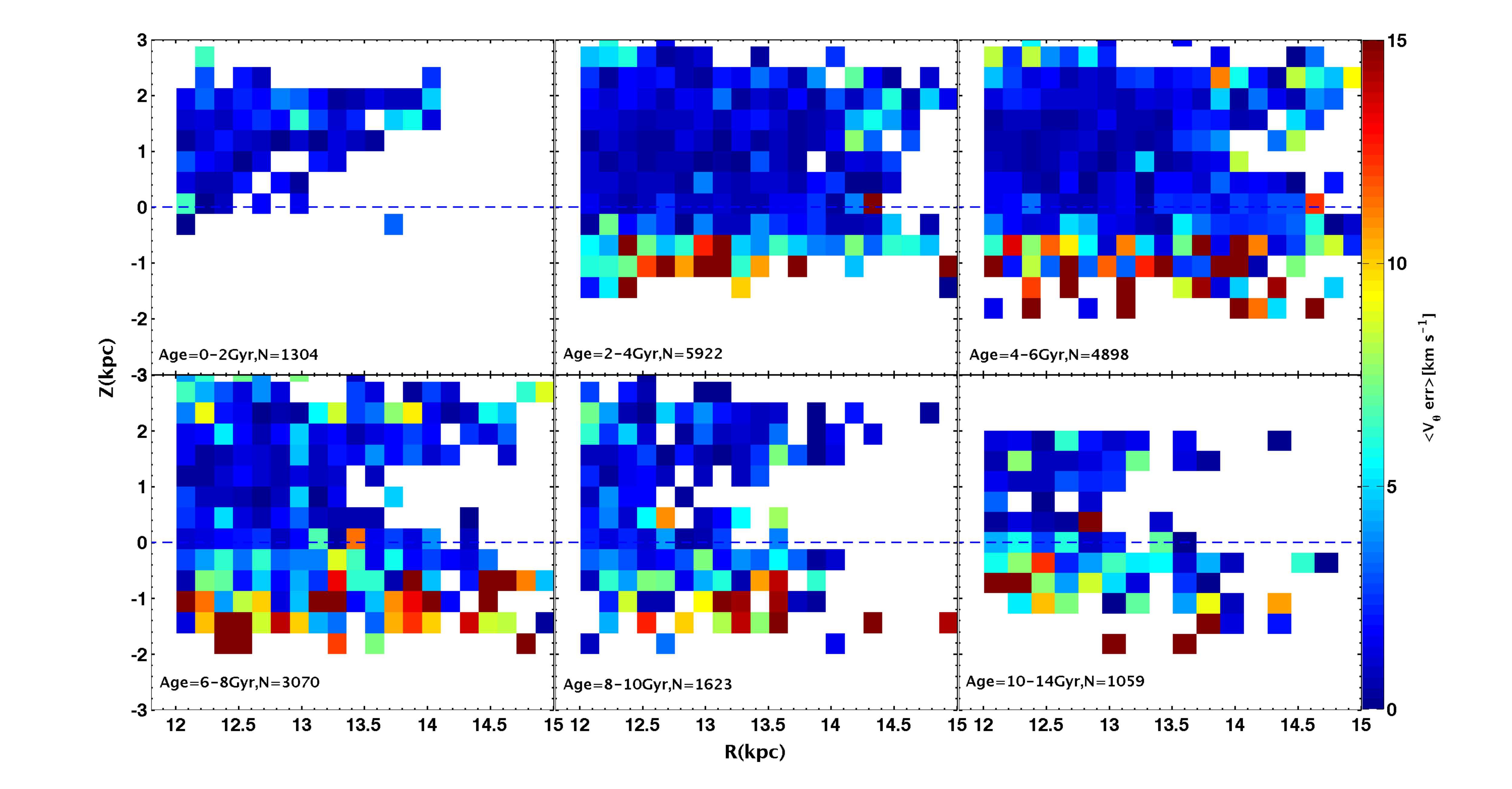}
  \caption{Similar to Fig. \ref{VRageerror}, but for the error analysis of the rotational ($V_\theta$) asymmetrical structures in the R, Z plane of the LAMOST-Gaia stars in different age populations. Most of the bins value is small with blue color, accompanied by some redder bins belonging to 10$-$15 km s$^{-1}$.}
  \label{Vthetaageerror}
\end{figure*}

We show the rotational velocity $V_\theta$ in different age populations in Fig. \ref{Vthetaage}. There is a clear asymmetry, with redder bins in the north compared to southern stars with bluer bins (i.e., slower rotational velocities), especially for heights larger than 0.4 \,kpc, the pattern is similar with the middle panel of Fig.~\ref{VRVPHIVZ}. The asymmetry exists in almost all bins, which is consistent with the radial velocity distribution, implying this structure is sensitive to possible perturbations over roughly the entire life of the Milky Way.  Let us focus on the general trend or average value of the symmetrical region in Fig. \ref{Vthetaage}, we can see the north stars are larger or stronger than south stars, most of north stars value can reach 245-250 km s$^{-1}$ , so many south stars are lower than this value, the absolute difference can reach 30 km s$^{-1}$  for many symmetrical bins, although there are fewer stars in the last two panels. we also do the error analysis for the rotational velocity in Fig. \ref{Vthetaageerror} , we find that many bins are around 0-5 km s$^{-1}$ , and some of these can reach around 15 km s$^{-1}$ , compared with velocity absolute difference mentioned before, we don't think it will change the asymmetrical pattern, and in the future, with the larger sample  and better age precision, we guess these patterns will be clearer.

\subsection{Discussion}

According to the selection of spectra and Hess diagram of the stars with photometric measurements in SDSS, \citet{xu2015} found and discussed the famous south middle structure which is located 4$-6$ \,kpc from the Sun, and they also investigated this structure's velocity and metallicity distribution. \citet{Wang2018b} found a density structure which is located at R$ \approx $13~kpc and the latitude is around $-$20 degree. 
The position and its distance of $\sim5$~kpc from the Sun confirm that this feature corresponds to the south middle substructure discovered in \citet{xu2015}. More interestingly, in \citet{Wang2018b}, a dip was clearly detected in one dimensional and two dimensional residual analysis of K giant star counts, which was described as new substructure located on the opposite side of the south middle structure \citep{Wang2018b}. 
Because this ``south middle opposite'' feature is mainly contributed by stars located at intermediate Galactic latitudes, they suggested that the effect of interstellar extinction should be negligible. Therefore, it is likely a real density structure in the outer disc \citep{Wang2018b}. We firstly confirm this structure in velocity space, showing that it has outward motions and a significant difference from the south middle region, and it has larger rotational velocity with contribution of asymmetric  drift. We can also say the south middle structure has a different pattern from the northern ``south middle opposite" region. However, we do not see clear north-south differences in vertical velocity (Fig.~\ref{VRVPHIVZ}), so we suggest that this ``south middle opposite" asymmetry is mainly caused by some mechanisms related to in$-$plane asymmetry.

Some possible mechanisms such as the Galactic bar's outer Lindblad resonance, perturbations due to the bar or spiral arms \citep {Denhen00,Fux01,Quillen05}, by external minor mergers such as the Sagittarius dwarf galaxy passing by \citep{Carrillo2019, Laporte18}, or by interaction with the Magellanic Clouds \citep{Gomez121, Gomez122, Minchev09, Minchev10} are proposed for these in-plane non-axisymmetries. In a future work, we will use toy models with the help of 
Galpy \citep{Bovy2015}, torus mapper with analytical method constructed from Hamiltonian mechanics action 
variables \citep{Binney2016}, and AGAMA action based dynamical method \citep{Vasiliev2019} to further investigate the dynamical origin in the outer disk.

\citet{Lopez2019} also produced kinematic maps using Gaia up to $R=20$ kpc. We find that we have the same pattern
with that work for the overall trend in the longitude=[120, 200]. There is a discrepancy in the range [-20, 20] and  [200, 240] attributed to LAMOST has low sampling rates in these regions.

\section{Conclusion} 

In this paper, using LAMOST$-$Gaia combined stars, we corroborate the existence of an asymmetrical structure in the  $V_R$ and $V_{\theta}$ distributions found recently in density distribution[R $\sim$ 12-15 \,kpc, Z $\sim$ 1.5 \,kpc] by \citet{Wang2018b}. We further investigate the variations of the asymmetrical pattern with different stellar ages, and find that the velocity asymmetries are present in samples from $\tau=0$\,Gyr to  $\tau=14$\,Gyr  for both $V_R$ and $V_{\theta}$. Both $V_R$ and $V_{\theta}$ in phase space have the same north-south asymmetry in the different stellar age bins, which suggests that the structure is real and different stellar populations respond to the perturbation by different levels of sensitivity. The stars with $\tau>8.0$ Gyr have a weaker pattern for $V_R$  in the phase space, probably because the old stars are kinematically hot, and do not respond to the perturbation as sensitively as young stars. However, we do not detect here corresponding asymmetry in the vertical motions; we only find a clear bending mode motion for the $V_Z$ distribution, which may be partly at odds with other works \citep[\S 4.3]{Lopez2019}. 

According to the features of the observed asymmetry, we infer that the possible perturbation to the disk asymmetrical structure mainly originated from an in-plane asymmetry mechanism, and is not likely due to scenarios which are related to the vertical motion. So we find that the in-plane asymmetry is decoupled with the vertical asymmetry for this range. We will explore these features in more detail in our future work.

 \acknowledgements
We would like to thank the anonymous referee for his/her helpful comments. We also thank Lawrence M. Widrow, Heidi J. Newberg for helpful discussions and comments. This work is supported by the National Natural Science Foundation of China 11833006, U1531244 and 11473001. MLC was supported by grant AYA2015-66506-P of the Spanish Ministry of Economy and Competitiveness. HFW is supported by the LAMOST Fellow project, National Key Basic Research Program of China via SQ2019YFA040021 and funded by China Postdoctoral Science Foundation via grant 2019M653504. JLC acknowledges support from the U.S. National Science Foundation via grant AST-1816196. The Guo Shou Jing Telescope (the Large Sky Area Multi-Object Fiber Spectroscopic Telescope, LAMOST) is a National Major Scientific Project built by the Chinese Academy of Sciences. Funding for the project has been provided by the National Development and Reform Commission. LAMOST is operated and managed by National Astronomical Observatories, Chinese Academy of Sciences. This work has also made use of data from the European Space Agency (ESA) mission {\it Gaia} (\url{https://www.cosmos.esa.int/gaia}), processed by the {\it Gaia} Data Processing and Analysis Consortium (DPAC, \url{https://www.cosmos.esa.int/web/gaia/dpac/consortium}). Funding for the DPAC has been provided by national institutions, in particular the institutions
participating in the {\it Gaia} Multilateral Agreement.


\begin{thebibliography}{}
\bibitem[Antoja et al.(2018)]{antoja2018} Antoja, T., Helmi, A., Romero-Gomez, M., et al. \ 2018, arXiv:1804.10196. (A18)
\bibitem[Bailer-Jones et al.(2018)]{Bailer-Jones2018} Bailer-Jones, C. A. L, Rybizki, J., Fouesneau, M., et al.\ 2018, arXiv:1804.10121B
\bibitem[Binney et al.(2016)]{Binney2016}Binney J., McMillan P. J., 2016, MNRAS, 456, 1982
\bibitem[Blanton et al.(2017)]{Blanton17} Blanton M. R., et al., 2017, AJ, 154, 28
\bibitem[Bovy et al.(2015)]{Bovy2015}Bovy J., 2015, ApJS, 216, 29
\bibitem[Chequers et al.(2018)]{Chequers18}Chequers,M. H., Widrow, L.M., Darling, K., 2018, MNRAS, 480, 4244
\bibitem[Cui et~al. (2012)]{cui2012}{Cui}, X.-Q., {Zhao}, Y.-H., {Chu}, Y.-Q., {et~al.} 2012, RAA, 12, 1197
\bibitem[Carrillo et al.(2019)]{Carrillo2019} Carrillo I., et al., 2019, submitted 
\bibitem[Denhen(2000)]{Denhen00}Dehnen, W., 2000, AJ, 119, 800
\bibitem[Deng et al.(2012)]{Deng2012}Deng, L. C., Newberg, H. J., Liu, C., et al. 2012, Research in Astronomy and Astrophysics, 12, 735
\bibitem[Fux et al.(2001)]{Fux01}Fux, R., 2001, A$\&$A, 373, 511 
\bibitem[Gaia Collaboration et~al. (2016)]{Gaia2016}Gaia Collaboration, Prusti, T., de Bruijne, J. H. J., et al. 2016, A\&A, 595, A1 
\bibitem[Gaia Collaboration et~al. (2018)]{Gaia2018}Gaia Collaboration, Brown, A. G. A., Vallenari, A., et al. 2018, A\&A, 616, A1.
\bibitem[Gaia Collaboration: Katz et~al. (2018)]{Katz2018}{Gaia Collaboration:} {Katz et al.}, D., {Antoja}, T., {Romero-G\'omez}, M., {et~al.} 2018, A\&A, Accepted (arXiv:1804.09380v1)
\bibitem[G{\'o}mez, Minchev \& Villalobos(2012)]{Gomez121}G{\'o}mez, F. A., Minchev, I., \& Villalobos, A. M. E. K., 2012, MNRAS, 419, 2163
\bibitem[G{\'o}mez et al.(2012)]{Gomez122}G{\'o}mez, F. A., et al., 2012, MNRAS, 423, 3727
\bibitem[Huang et al.(2015)]{Huang2015}Huang Y., Liu X.-W., Yuan H.-B., Xiang, M.-S., Huo Z.-Y., Chen B.-Q., Zhang Y., Hou Y.-Y., 2015, MNRAS, 449, 162
\bibitem[Huang et al.(2019)]{Huang2019} Huang et al., 2019,  to be submitted to ApJ
\bibitem[Liu et al.(2014)]{liu2014} Liu, X.~W., Yuan, H.~B., Huo, Z.~Y., et al.\ 2014, IAUS, 298, 310L
\bibitem[\protect\citeauthoryear{Laporte et al.}{2018}]{Laporte18} Laporte C.~F.~P., G{\'o}mez F.~A., Besla G., Johnston K.~V., Garavito-Camargo N., 2018, MNRAS, 473, 1218 
\bibitem[L\'opez-Corredoira \& Sylos Labini(2019)]{Lopez2019}L\'opez-Corredoira, M., \& Sylos Labini, F.\ 2019, A$\&$A, 621, A48 
\bibitem[Minchev et al.(2009)]{Minchev09}  Minchev, I., Quillen, A. C., Williams, M., Freeman, K. C., Nordhaus, J., Siebert, A., \& Bienaym\'{e}, O., 2009, MNRAS, 396, L56
\bibitem[Minchev et al.(2010)]{Minchev10}Minchev, I., Boily, C., Siebert, A., Bienayme, O., 2010, MNRAS, 407, 2122 
\bibitem[Quillen \& Minchev(2005)]{Quillen05}Quillen, A. C., \& Minchev, I., 2005, AJ, 130, 576
\bibitem[Sch\"onrich et al.(2012)]{Schonrich12} Sch\"onrich, R., et al., 2012, MNRAS, 427, 274
\bibitem[Tian et al.(2015)]{Tian15}Tian, H. J., Liu, C., Carlin, J. L., et al. 2015, ApJ, 809, 145
\bibitem[Vasiliev et al.(2019)]{Vasiliev2019}Vasiliev E., 2019, MNRAS, 482, 1525V
\bibitem[Wang et al.(2018a)]{Wang2018a}Wang, H. F., L\'opez-Corredoira, M., Carlin, J. L., et al., 2018a, MNRAS, 477, 2858
\bibitem[Wang et al.(2018b)]{Wang2018b}  Wang, H. F., Liu, C., Xu, Y., et al. 2018b, MNRAS, 478, 3367
\bibitem[Wang et al.(2018c)]{Wang2018c}Wang, H. F., Liu, C., Deng, L. C., 2018c, IAU Symposium 334: Rediscovering our Galaxy, eds. C. Chiappini, I. Minchev, E. Starkenberg, M. Valentini, 378
\bibitem[Wang et al.(2019b)]{wang2019b} Wang, H.~F., {et~al.} 2019b, MNRAS, submitted 
\bibitem[Wang et al.(2019c)]{wang2019c} Wang, H.~F., {et~al.} 2019c, IAU Symposium 353: Galactic Dynamics in the Era of Large Surveys, eds. J. Shen, M. Valluri, \& J. A. Sellwood, in press
\bibitem[Widrow et al.(2012)]{Widrow12}Widrow, L. M., Gardner, S., Yanny B., Dodelson S., $\&$ Chen H.-Y. 2012, ApJ, 750, L41
\bibitem[Widrow et al.(2014)]{Widrow14}Widrow, L. M., Barber, J., Chequers, M. H., \& Cheng, E., 2014, MNRAS, 440, 1971
\bibitem[Xu et al.(2015)]{xu2015} Xu, Y., Newberg, H.~J., Carlin, J.~L., et al.\ 2015, \apj, 801, 105 
\bibitem[Zhao et~al.(2012)]{zhao2012}{Zhao}, G., {Zhao}, Y.~H., {Chu}, Y.~Q., {et~al.} 2012, RAA, 12, 723

\end{thebibliography}
\end{document}